\documentclass{article}

\usepackage[final]{neurips_2023}

    \usepackage[final]{neurips_2023}

\usepackage[utf8]{inputenc} 
\usepackage[T1]{fontenc}    
\usepackage{hyperref}       
\usepackage{graphicx}
\usepackage{tikz}
\usepackage{psfrag}
\usepackage{forest}
\usepackage{makecell}
\usepackage{adjustbox}
\usepackage{booktabs}
\usepackage{multirow}
\usepackage{enumitem}

\usepackage{url}            
\usepackage{booktabs}       
\usepackage{amsfonts}       
\usepackage{nicefrac}       
\usepackage{microtype}      
\usepackage{xcolor}         
\usepackage{color}
\usepackage{amsmath}

\title{LLM4EDA: Emerging Progress in Large Language Models for Electronic Design Automation}

\author{%
Ruizhe Zhong$^{1}$\thanks{Contributed Equally.} \And Xingbo Du$^{1*}$ \And Shixiong Kai$^{2}$ \And Zhentao Tang$^{2}$ \And Siyuan Xu$^{2}$ \And Hui-Ling Zhen$^{2}$ \And Jianye Hao$^{2,3}$ \And Qiang Xu$^{4}$ \And Mingxuan Yuan$^{2}$ \And  Junchi Yan$^{1}$\thanks{Correspondence author.} \\
\\
$^1$ Dept. of Computer Science and Engineering \& MoE Key Lab of AI, Shanghai Jiao Tong University \\ 
\texttt{\{zerzerzerz271828, duxingbo, yanjunchi\}@sjtu.edu.cn} \\
\\
$^2$ Huawei Noah’s Ark Lab \\ 
\texttt{\{kaishixiong, tangzhentao1, xusiyuan520\}@huawei.com} \\
\texttt{\{zhenhuiling2, haojianye, yuan.mingxuan\}@huawei.com} \\
\\
$^3$ College of Intelligence and Computing, Tianjin University \\
\\
$^4$ Department of Computer Science and Engineering, The Chinese University of Hong Kong \\
\texttt{qxu@cse.cuhk.edu.hk}
}

\begin{document}

\maketitle

 
\setcounter{footnote}{0}

\begin{abstract}
Driven by Moore's Law, the complexity and scale of modern chip design are increasing rapidly. Electronic Design Automation (EDA) has been widely applied to address the challenges encountered in the full chip design process. 
However, the evolution of very large-scale integrated circuits has made chip design time-consuming and resource-intensive, requiring substantial prior expert knowledge. 
Additionally, intermediate human control activities are crucial for seeking optimal solutions. 
In system design stage, circuits are usually represented with Hardware Description Language (HDL) as a textual format.
Recently, Large Language Models (LLMs) have demonstrated their capability in context understanding, logic reasoning and answer generation. 
Since circuit can be represented with HDL in a textual format, it is reasonable to question whether LLMs can be leveraged in the EDA field to achieve fully automated chip design and generate circuits with improved power, performance, and area (PPA). 
In this paper, we present a systematic study on the application of LLMs in the EDA field, categorizing it into the following cases: 
1) assistant chatbot, 
2) HDL and script generation, 
and 3) HDL verification and analysis. 
Additionally, we highlight the future research direction, focusing on applying LLMs in logic synthesis, physical design, multi-modal feature extraction and alignment of circuits.
We collect relevant papers up-to-date in this field via the following link: 
\url{https://github.com/Thinklab-SJTU/Awesome-LLM4EDA}.
\end{abstract}

\section{Introduction}
Over the past few decades, Electronic Design Automation (EDA) algorithms and tools have made significant strides, yielding substantial improvements in chip design productivity. 
At the same time, driven by Moore's Law, circuit sizes have exponentially increased, presenting new challenges for chip engineers in achieving Very Large-Scale Integration (VLSI) with billions of transistors. 
In addition to the scale, it has become increasingly challenging to satisfy the demands of Power, Performance, and Area (PPA), specifications, and other constraints, especially throughout the entire and long EDA design flow.
During this long design flow, the involvement of numerous intermediate processes necessitates time- and cost-intensive human intervention, often requiring iterative interactions with natural language or programming language interfaces.
These processes generate abundant and various outputs and logs rich in textual information, demanding engineers' understanding, processing, and decision-making for following guidance and commands. 
Consequently, the complete design flow remains far from being fully automated.
Simultaneously, chip design also imposes high demands on engineers, and it typically takes several years to cultivate an experienced engineering professional in this field.
How to achieve full automation of the circuit design process and reduce reliance on experienced circuit design engineers has become a key focus of research.

Deep learning, an ever-advancing technology, has found widespread application across diverse domains and scenarios, including classification, detection, forecasting, and generation. 
Notably, it exhibits great potential in generating high-quality solutions for many NP-complete (NPC) problems, which commonly arise in the EDA field. 
Traditional methods encounter challenges in effectively addressing these problems due to their demanding computational resources and time requirements, particularly in the realm of VLSI. 
Unlike traditional approaches that tackle each problem independently without the accumulation of knowledge, deep learning methods excel at extracting high-level features and representations shared among similar or related cases. 
Leveraging these features allows for their reuse and application throughout the problem-solving process, resulting in enhanced speed and improved solution quality. 
Consequently, the integration of deep learning techniques to aid and accelerate the resolution of EDA problems represents a highly promising direction of research.

At present, deep learning has found extensive applications through flow of chip design, including logic synthesis~\citep{hosny2020drills,yuan2023easyso}, floorplanning~\citep{amini2022generalizable}, placement~\citep{lin2019dreamplace, mirhoseini2021graph, cheng2021joint, cheng2022policy, lai2022maskplace}, clock tree synthesis~\citep{lu2021clock, liang2023bufformer} routing~\citep{cheng2022policy, du2023hubrouter} and 
PPA prediction~\citep{guo2022timing, chai2023circuitnet,zhong2024preroutgnn}. 
Simultaneously, many works focus on general representation learning of circuits~\citep{li2022deepgate, shi2023deepgate2, wang2022functionality}.
It embeds both functionality and structural information of a circuit as vectors, and these representations can be further utilized in various downstream tasks rather than learning a specific model for each task from scratch.
To support the demand of neural network training for massive training data and achieve stronger generalization in EDA, open-sourced datasets such as Circuit~\citep{chai2023circuitnet} and Circuit 2.0~\citep{anonymous2023circuitnet} have been made available to the research community. These datasets offer cross-stage and cross-design samples, facilitating comprehensive exploration of advancements in Artificial Intelligence for Electronic Design Automation (AI4EDA).
We have collected papers and maintained real-time updates about AI4EDA~\footnote{\url{https://github.com/Thinklab-SJTU/awesome-ai4eda}}.

Recently, Large Language Models (LLMs) demonstrated capability in various aspects, including context understand, Question and Answer (Q\&A) and logic reasoning.
Both commercial (ChatGPT, Bard, etc.) and open-source (LLaMA2~\citep{touvron2023llama}) schemes have achieved significant advancement in this field.
Since circuits can be depicted in programming language, Hardware Description Language (HDL), specifications and many intermediate outputs through EDA flow are also represented in text format. It is natural to ask whether LLMs can be leveraged for EDA to assist engineers in chip design.

\begin{figure}[!tb]
    \centering
    \resizebox{0.88\linewidth}{!}{
    \includegraphics[width=\textwidth]{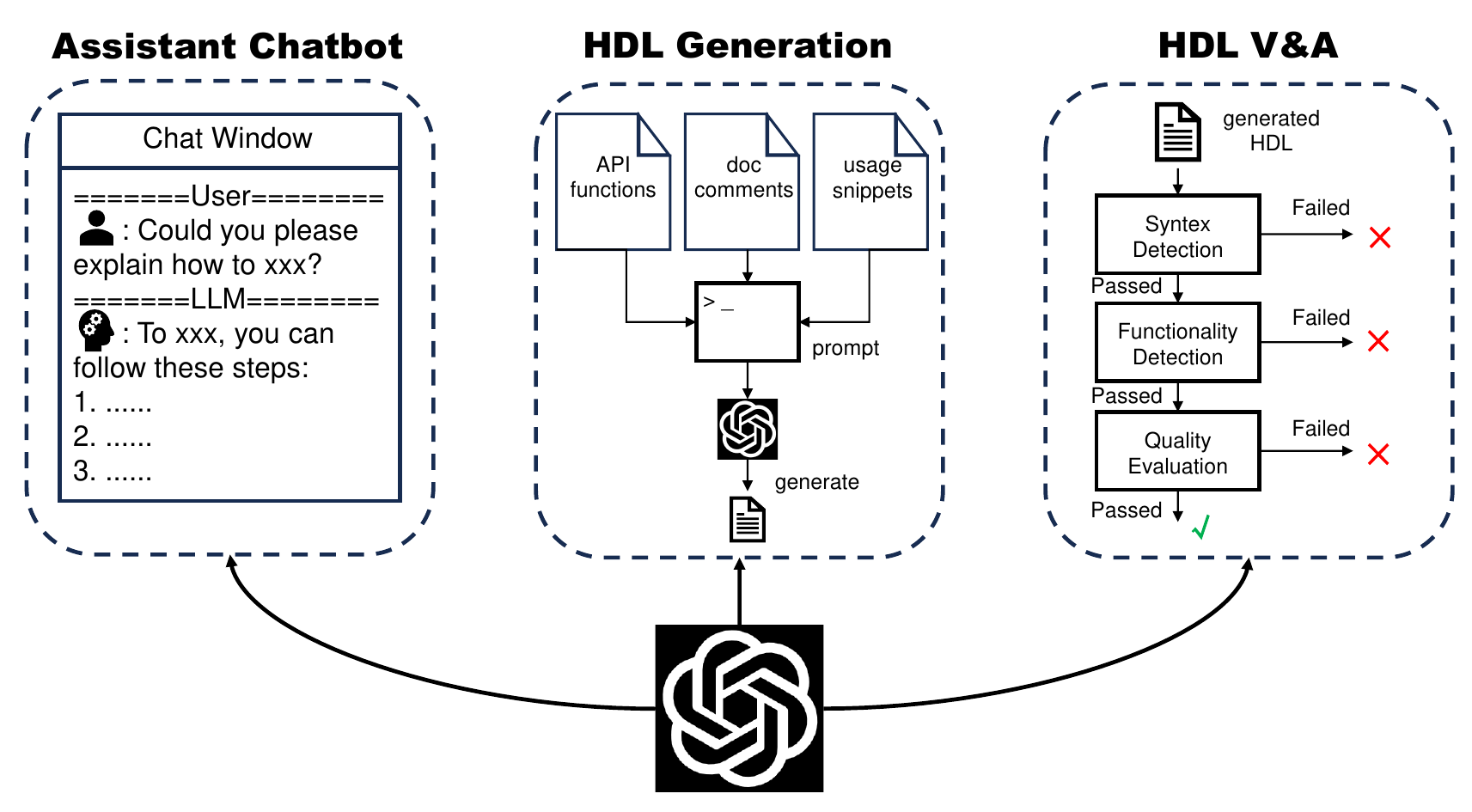}
    }
    \caption{Three categories in terms of progress and application in Large Language Models for Electronic Design Automation (LLM4EDA).}
    \label{fig:task_sketch_shrink}
\end{figure}

In this survey, we conduct a comprehensive and detailed investigation on the progress and application in Large Language Models for Electronic Design Automation (LLM4EDA). As shown in Fig.~\ref{fig:task_sketch_shrink}, we categorize current application and exploration of LLMs in EDA into the following three directions:
\begin{itemize}[leftmargin=0.175in]
    \item \textbf{Assistant Chatbot}. Users can interact with LLMs for knowledge acquisition and Q\&A, without spending time in waiting or actively searching for answers.
    We present some notable works which aim to provide user-friendly and easy-interactively assistant chatbot and bring us new interaction paradigm with EDA software.
    
    \item \textbf{HDL and Script Generation}. Given language format specification and requirements, LLMs will generate RTL codes and EDA controlling scripts.
    This automated process streamlines the development of circuit designs by reducing the manual effort traditionally involved in writing codes and scripts, thereby enhancing efficiency and productivity in the chip design. 
    Besides, how to evaluate the quality of generated codes remains an open research focus, where syntax correctness and functionality equivalence are key factors.
    In EDA field there exist more metrics to consider, such as Power, Performance and Area (PPA) and security issues.
    Therefore, new evaluation framework considering various aspects are also essential.
    

    \item \textbf{HDL Verification and Analysis}.
    We also investigate LLMs' wide application in code analysis, such as bug detecting \& fixing, code summarization and security checking.
    Besides, LLMs have also demonstrated strong ability for verification, e.g. Assertion Based Verification.
    
\end{itemize}

Apart from the aforementioned advancements, other key processes in EDA flow also exhibit promising potential when viewed from an LLM perspective. Specifically, we provide following overviews:

\begin{itemize}[leftmargin=0.175in]
    \item \textbf{Logic Sythesis}. 
    LLM holds the potential to generate an optimization sequence, along with the corresponding arguments in logic synthesis, when Hardware Description Language (HDL) and prompts are considered as inputs.
    \item \textbf{Physical Design}. The complexity of Placement and Routing (P\&R) currently makes it challenging to directly apply LLMs. However, employing strategies such as graph partitioning and clustering can reduce the scale of the problem and accelerate the solving process, thereby making the use of LLMs feasible. LLMs could also be used to optimize modules or act as a reward system for reinforcement learning agents.
\end{itemize}


\section{Preliminaries}
\subsection{Large Language Models}
Large Language Models (LLMs)~\citep{openai2023gpt4, touvron2023llama} are a type of artificial intelligence models characterized by their vast number of parameters. These models are trained on substantial amounts of text data over extended GPU time. The pioneer in this field is the GPT series~\citep{brown2020language, NEURIPS2022_b1efde53, openai2023gpt4} from OpenAI, where GPT-3~\citep{brown2020language} is an autoregressive language model with 175 billion parameters, significantly outperforming other contemporary models in terms of scale. On this basis, GPT-3.5~\citep{NEURIPS2022_b1efde53} focuses on the fine-tuning GPT-3, particularly incorporating the reinforcement learning from human feedback (RLHF)~\citep{christiano2017deep, stiennon2020learning} to enhance alignment with human preferences. While these models demonstrate impressive performance, the official surge in LLMs can be attributed to the advent of ChatGPT~\footnote{\url{https://openai.com/blog/chatgpt/}}, which adapts GPT-3.5 to dialogue, achieving remarkable results and applications. In response to the widespread interest in ChatGPT, OpenAI developed GPT-4~\citep{openai2023gpt4}, a more capable LLM than ChatGPT, that supports multi-modal, longer, and more logical input/output.

The success of the GPT series has spurred the development of various other LLMs, such as LLaMA~\citep{touvron2023llama} and Gemini~\footnote{\url{https://deepmind.google/technologies/gemini/}} by different corporations and institutes. Empowered by extensive data, a large number of parameters, and prolonged training time, LLMs can generate human-like text given certain inputs. They are not only capable of performing tasks like translation, character simulation, and generating creative content like poems and stories, but also answering questions and handling high-difficulty tasks such as generating executable code and scripts. The potential of LLMs is vast and far from being fully realized.

\subsection{Design Flow of Electronic Design Automation}
\begin{figure}[!tb]
    \centering
    \includegraphics[width=\textwidth]{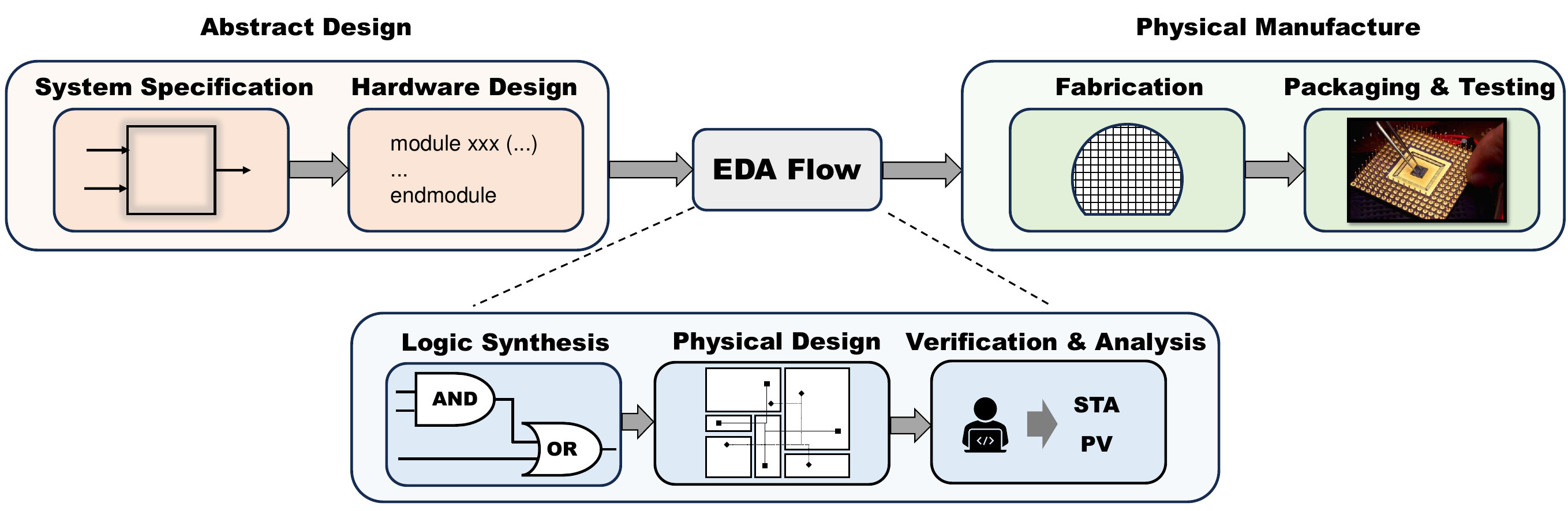}
    \caption{The typical flow of digital chip design.}
    \label{fig:chip_design_flow}
    \vspace{-10pt}
\end{figure}

\begin{figure}[!tb]
    \centering
    \includegraphics[width=\textwidth]{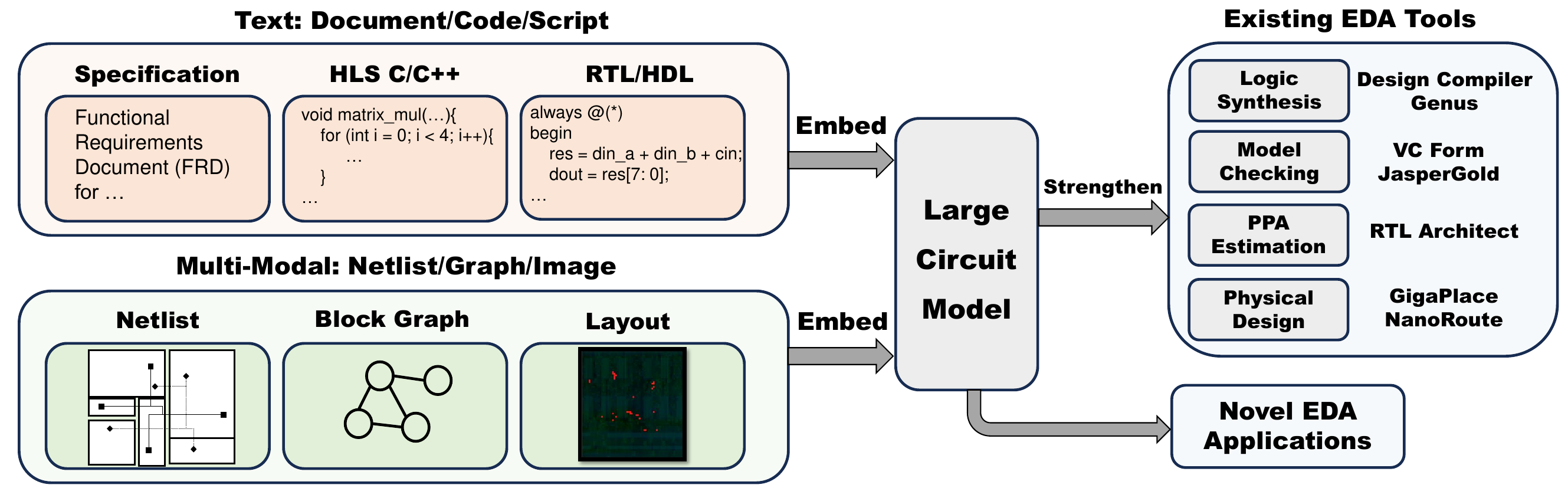}
    \caption{The flow driven by large circuit model.}
    \label{fig:large_circuit_model}
    \vspace{-10pt}
\end{figure}

The process of chip design is intricate and multifaceted, typically segmented into several distinct stages~\citep{anonymous2023circuitnet}. As shown in Fig.~\ref{fig:chip_design_flow}, these stages encompass abstract design, EDA flow, and physical manufacturing. The EDA flow itself can be further subdivided into logic synthesis, physical design, verification, and analysis, each of which plays a crucial role in the successful creation of a chip. We also envision that a large circuit model, regarded as an EDA-guided LLM, could union different stages and output a general solution in the EDA flow. Specifically, our approach utilizes a data-driven large model, where both text and multi-modal information are integrated into a latent space. The text data encompasses specification documents, HLS codes, and RTL scripts, while the multi-modal information comprises netlists, graphs, and images. The large circuit model is trained either from scratch or by fine-tuning an existing Language Large Model (LLM) using these data. Once trained, this model can enhance the capabilities of existing EDA tools and facilitate various downstream applications. The ideal workflow, based on the large circuit model, is depicted in Fig.~\ref{fig:large_circuit_model}. However, it’s important to note that achieving our ultimate goal is still a considerable journey ahead.

\section{Application of LLM4EDA}
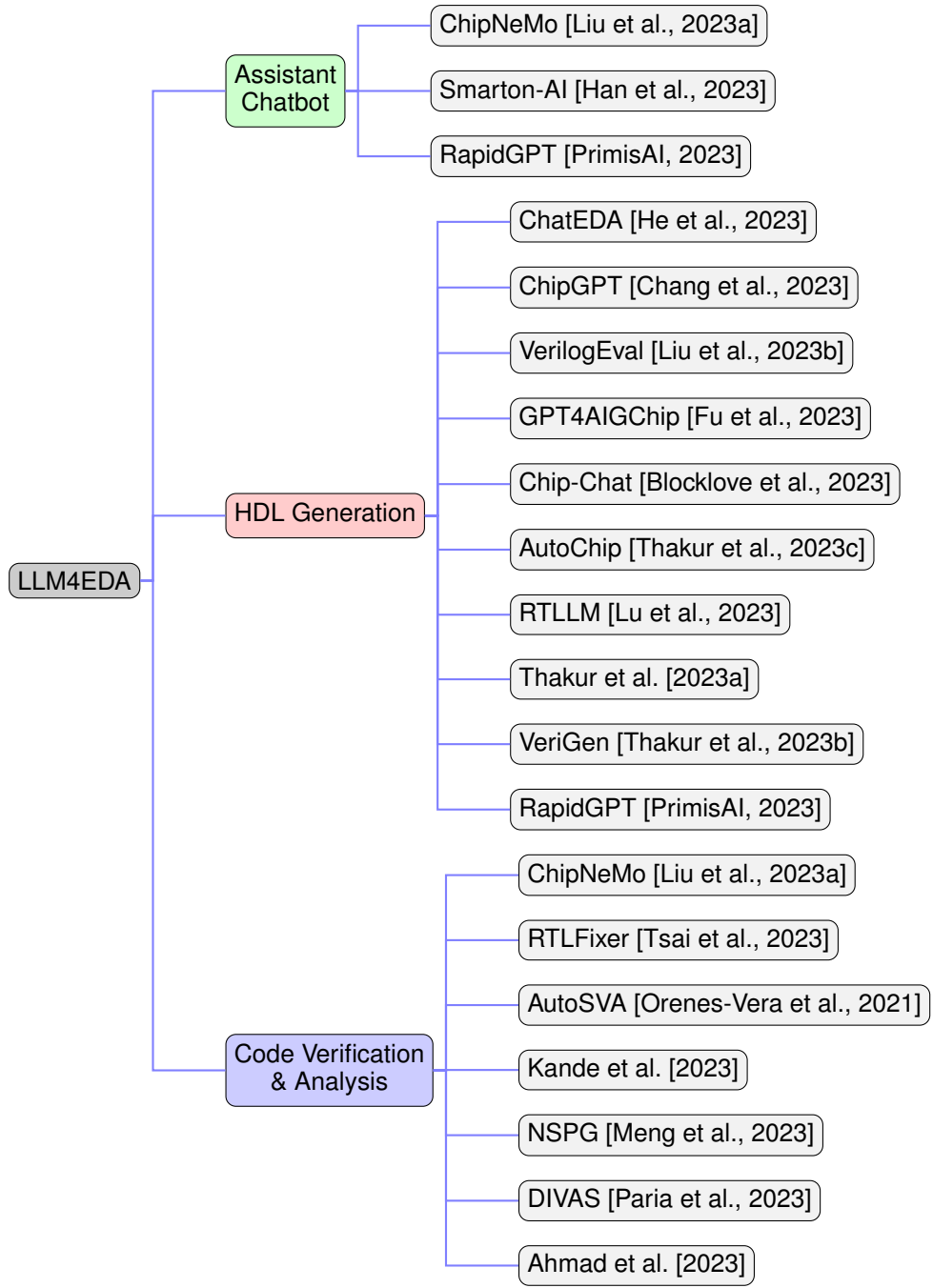
\begin{figure}
    \centering
{
    \begin{forest}
for tree={
    grow'=east,
    draw,
    rounded corners,
    node options={font=\sffamily, fill=blue!20},
    edge={thick, color=blue!50},
    l sep+=0.8cm,
    s sep+=0.1cm,
    parent anchor=east,
    child anchor=west,
    anchor=west,
    edge path={
        \noexpand\path [draw, \forestoption{edge}] (!u.parent anchor) -- +(5pt,0) |- (.child anchor)\forestoption{edge label};
    },
}
[LLM4EDA, fill=black!20
    [\makecell{Assistant \\ Chatbot}, for tree={fill=green!20}
        [ChipNeMo~\citep{liu2023chipnemo}, for tree={fill=black!5}]
        [Smarton-AI~\citep{han2023new}, for tree={fill=black!5}]
        [RapidGPT~\citep{RapidGPT}, for tree={fill=black!5}]
    ]
    [\makecell{HDL Generation}, for tree={fill=red!20}
            [ChatEDA~\citep{he2023chateda}, for tree={fill=black!5}]
            [ChipGPT~\citep{chang2023chipgpt}, for tree={fill=black!5}]
            [VerilogEval~\citep{liu2023verilogeval}, for tree={fill=black!5}]
            [GPT4AIGChip~\citep{fu2023gpt4aigchip}, for tree={fill=black!5}]
            [Chip-Chat~\citep{blocklove2023chip}, for tree={fill=black!5}]
            [AutoChip~\citep{thakur2023autochip}, for tree={fill=black!5}]
            [RTLLM~\citep{lu2023rtllm}, for tree={fill=black!5}]
            [\cite{thakur2023benchmarking}, for tree={fill=black!5}]
            [VeriGen~\citep{thakur2023verigen}, for tree={fill=black!5}]
            [RapidGPT~\citep{RapidGPT}, for tree={fill=black!5}]
    ]
    [\makecell{Code Verification \\ \& Analysis}, for tree={fill=blue!20}
        [ChipNeMo~\citep{liu2023chipnemo}, for tree={fill=black!5}]
        [RTLFixer~\citep{tsai2023rtlfixer}, for tree={fill=black!5}]
        [AutoSVA~\citep{orenes2021autosva}, for tree={fill=black!5}]
        [\cite{kande2023llm}, for tree={fill=black!5}]
        [NSPG~\citep{meng2023unlocking}, for tree={fill=black!5}]
        [DIVAS~\citep{paria2023divas}, for tree={fill=black!5}]
        [\cite{ahmad2023fixing}, for tree={fill=black!5}]
    ]
]
\end{forest}
}
\caption{Research Tree of Large Language Models for Electronic Design Automation.}
\label{fig:tree}
\end{figure}

According to the mainstream research interests, we synthesize recent literature into four aspects, including 1) assistant chatbot; 2) HDL and script generation; 3) evaluation of generated code, and 4) code verification and analysis. Related papers are listed in Fig.~\ref{fig:tree}.

\subsection{Assistant Chatbot}
The full chip design flow necessitates extensive prior expert knowledge, which requires years of research and accumulation. To assist engineers in obtaining answers to their questions related to architecture, design, tools, and verification, an engineering assistant chatbot can effectively meet their needs.
Although current LLMs are pre-trained on various types of text and serve as general chatbots, they may lack profound and accurate understanding in specific domains like the EDA field.
Therefore, developing an engineering assistant chatbot that incorporates knowledge extracted from internal design documents, code, and recorded data pertaining to designs and technical communications could substantially enhance design productivity.

It is widely acknowledged that LLMs may generate inaccurate answers, giving the impression of correctness while often leading to a high level of misunderstanding. 
This phenomenon is commonly referred to as hallucination~\citep{ji2023survey}. 
While the exact causes of hallucination are not yet fully comprehended, it is imperative to address this issue, particularly in engineering, with a specific emphasis on the EDA field. 
The accuracy of these generated answers holds significant importance in engineering, underscoring the need to mitigate the occurrence of hallucination.

A representative work is ChipNeMo~\citep{liu2023chipnemo}. 
They propose to leverage retrieval augmented generation (RAG)~\citep{lewis2020retrieval} method to mitigate hallucination.
Also, they find that fine-tuning an off-the-shelf unsupervised pre-trained dense retrieval model with a modest amount of domain specific training data significantly improves retrieval accuracy.
They utilize these techniques to fine-tune a pre-trained LLM to construct a chip-design-specific assistant chatbot.
As a consequence, engineers can focus more on brainstorming, designing, and writing codes, instead of waiting answers or searching knowledge they lack.
RapidGPT~\citep{RapidGPT} is another notable work and it is the industry’s first AI-based pair-designer tailored for FPGA engineers.
Serving as an intelligent code assistant, RapidGPT leverages AI algorithms to provide accurate and context-aware code suggestions, allowing FPGA engineers to write Verilog code more efficiently.
RapidGPT also provides conversational capabilities to the next level by offering a chat panel, allowing users to easily communicate with the tool. 
This chat panel can be used to write or improve HDL code in a conversational manner.

From the perspective of interacting with LLMs, Smarton-AI~\citep{han2023new} propose a new interaction paradigm for leveraging LLMs in complex software, including main-sub GPTs and a Q\&A GPT.
The main GPT identifies most relevant tasks according to user input in natural language format.
All sub-GPTs form a Mixture of Experts (MoE)~\citep{masoudnia2014mixture} system. Each sub-GPT is dedicated to a identified task from main-GPT and generates learning documents for this task.
The Q\&A GPT is responsible for processing the generated learning documents produced by the sub-GPTs. It serves as an assistant chatbot, leveraging the input derived from these documents to facilitate user interactions and provide relevant information in a question-and-answer format.

\subsection{HDL Generation \& Evaluation}
\subsubsection{HDL Generation}
Modern chip design often originates from the specifications given by human natural language and then turns to translations into Hardware Description Languages (HDLs) such as Verilog. These translations typically require highly-skilled hardware engineers and suffer from man-made errors and loads of labors. Automating the hardware design could effectively reduce human errors and accelerate the process of design translations. Equipped with LLMs which are empirically proved to be effective to generate high-quality contexts, it is natural to apply LLMs to language translations and other related script generation. 

Recent works of LLMs for HDL and script generation mainly emphasize on two aspects: 1) which flow to work on, and 2) what the goal is.

Among the works, ChatEDA~\citep{he2023chateda} endeavors to automate the flow from Register-Transfer Level (RTL) to Graphic Data System Version II (GDSII), which splits the flow stream into task planning, script generation, and task execution due to the complexity of the whole flow. Specifically, ChatEDA treats the natural language as input and then generates effective codes for task execution with EDA tools. Similar flow splitting can also been observed in  ChipGPT~\citep{chang2023chipgpt} that reports a four-stage logic design framework, including generating prompts, producing initial Verilog programs, correcting and optimizing these programs, and selecting the optimal design according to the target metrics. GPT4AIGChip~\citep{fu2023gpt4aigchip} constructs a framework that intends to democratize AI accelerator design. By investigating the weaknesses of current LLMs, especially the inability to handle lengthy codes, GPT4AIGChip also decouples various hardware modules and functionalities of the AI accelerator design to enable the LLM-driven design automation. Chip-Chat~\citep{blocklove2023chip} targets the case for the novel 8-bit accumulator-based microprocessor architecture and partitions the task into generating Verilog code and producing most of the processor's specification.


To enable the generated code to be more functionally accurate, AutoChip~\citep{thakur2023autochip} focuses more on the context from compilation errors and dubugging contents when incorporating the interactions from LLMs into the output of the Verilog simulations. Unlike AutoChip~\citep{thakur2023autochip} and most other LLM-baesd approaches that targets the design correctness, RTLLM~\citep{lu2023rtllm}, as an instruction for generating design RTL, concerns more about the design qualities, which is fairer in practice. In addition to the metric-based goals, Chip-Chat~\citep{blocklove2023chip} seeks a von Neumann type design with 32 bytes of memory considering the space restriction.
For the security issues and vulnerabilities in hardware design, \cite{nair2023generating} constructs robust prompt for ChatGPT to generate design resistant Common Vulnerability Enumerations (CWEs).

VerilogEval~\citep{liu2023verilogeval} constructed a synthetic supervised fine-tuning dataset by leveraging GPT-3.5~\citep{NEURIPS2022_b1efde53} to generate problem descriptions paired with Verilog code.
RTLCoder~\citep{liu2023rtlcoder} follows this synthetic method, and proposes a new LLM fine-tuning algorithm leveraging quality score feedback.  Furthermore, they also quantize LLM to 4-bit with a total size of 4GB, enabling it to function on a single laptop with only slight performance degradation.
Simultaneously, many works focus on prompt engineering and feedback, directly utilizing existed general LLMs without further fine-tuning.
ChipGPT~\citep{chang2023chipgpt} uses template-based prompts, providing details and purpose of original specification. It also contains an output manager to provide LLM with PPA or other human-specified targets as feedback.
\cite{du2023power} proposes to utilize In-Context Learning and Chain-of-Thought prompting techniques in complex FPGA design to tackle challenges, including sub-task scheduling and multi-step thinking problems. 
Chip-Chat~\citep{blocklove2023chip} proposes the conversation flow technique, breaking large design into sub-tasks and giving output from previous sub-task to LLM as base specification and feedback.




\subsubsection{HDL Evaluation}
\label{sec:eval-gen-code}
Once the LLMs have been developed for code generation, it becomes essential to evaluate their quality. 
This evaluation process involves checking for both syntax correctness and functionality correctness. 
Syntax correctness ensures that the generated code follows the proper programming language syntax rules, while functionality correctness ensures that the code performs the intended tasks accurately.

In addition to these checks, when dealing with codes in the EDA field, it is crucial to conduct further testing to evaluate their final power, performance, and area (PPA) characteristics. This testing process typically takes place after the completion of the entire design flow. They are utilized to verify whether the final circuit design meets the specified requirements and corresponding constraints, such as area utilization and timing constraints.

Take RTLLM~\citep{lu2023rtllm} as instance: it provides an public benchmark to evaluate the generated codes from following three perspectives:
\begin{enumerate}[leftmargin=0.175in]
    \item Syntax goal. It means that the generated RTL design should as least be correct, which can be verified by checking whether the design can correctly synthesized into netlist.
    \item Functionality goal. It means the functionality of generated RTL should be exactly the same as user's expectation. This goal can be checked with a comprehensive testbench.
    \item Design quality goal. If the generated RTL pass both above-mentioned unit tests, we need to further check its design quality, e.g. PPA. It can be verified by checking the PPA values after the synthesis and layout of RTL.
\end{enumerate}

\begin{table}[!tb]
    \centering
    \caption{Different datasets for fine-tuning towards LLM4EDA.
    Chat: assistant chatbot. Gen.: HDL and script generation. V\&A: HDL verification and analysis.}
    \adjustbox{width=1.0\linewidth}{
    \begin{tabular}{c|ccc|c|c}
        \toprule
         \multirow{2}{*}{\textbf{Dataset}} & \multicolumn{3}{c|}{\textbf{Task}} & \multirow{2}{*}{\textbf{Size}} &  {\centering \multirow{2}{*}{\textbf{Description}}}  \\
         ~ & \textbf{Chat} & \textbf{Gen.} & \textbf{V\&A} & ~ & ~ \\

         \midrule\midrule
         
        \makecell{ChipNeMo \\ \citep{liu2023chipnemo}} & $\checkmark$ & $\checkmark$ & $\checkmark$ & 24.1 B tokens & \multicolumn{1}{m{8cm}}{Data from NVBugs (NVIDIA’s internal bug database), bug summary, design source, documentation, verification. LLaMA2 tokenizer is adapted and approximately 9K new tokens are added to improve tokenization efficiency.} \\
        
        \midrule
        \makecell{ChatEDA \\ \citep{he2023chateda}} & ~ & \checkmark & & 1,500 instructions & \multicolumn{1}{m{8cm}}{Instruction tuning: Query GPT-4 to generate and  collect instructions. The core controller, AutoMage is further fine-tuned on these instructions.} \\
        
        \midrule
        \makecell{GPT4AIGChip \\ \citep{fu2023gpt4aigchip}} & ~ & \checkmark & ~ & 7,000 snippets & \multicolumn{1}{m{8cm}}{Open-source HLS code snippets from GitHub  and customized HLS templates with implementation  instructions to fine-tune LLMs.} \\
        
        \midrule
        \makecell{VeriGen \\ \citep{thakur2023verigen}} & ~ & \checkmark & ~ & 400 MB & \multicolumn{1}{m{8cm}}{From Verilog textbooks and  open-source GitHub repositories. Training samples are further generated through overlapped sliding windows on module blocks.}\\
        
        \midrule
        \makecell{\cite{dehaerne2023deep}} & ~ & \checkmark & ~ & 100,000 files & \multicolumn{1}{m{8cm}}{Verilog and SystemVerilog from  GitHub open-source repositories.  The dataset consists of two unlabeled subsets,  file-level data and snippet-level data, and  a labeled subset of snippet definition and body pairs}\\
        
        \midrule
        \makecell{VerilogEval \\ \citep{liu2023verilogeval}} & ~ & \checkmark & ~ & 8,502 samples & \multicolumn{1}{m{8cm}}{Designs generated from GPT-3.5 for SFT data, containing description and corresponding code. MinHash algorithm with Jaccard similarity is also performed to realize approximate deduplication.}\\
        
        \midrule
        \makecell{RTLCoder \\ \citep{liu2023rtlcoder}} & ~ & \checkmark & ~ & 10,000 designs & \multicolumn{1}{m{8cm}}{Generated from GPT-3.5, each sample consists of an description and corresponding RTL code. Conditional log probability based quality score  is also incorporated for fine-tuning.}\\
        
        \midrule
        \makecell{NSPG \\ \citep{meng2023unlocking}} & ~ & ~ & \checkmark & 20,000 sentences & \multicolumn{1}{m{8cm}}{Documentation collected from OpenTitan, RISC-V,   OpenRISC, MIPS, OpenSPARC. This dataset is further  augmented with Random Deletion, Random Swap, Synonym Replacement and Random Insertion.} \\
        
         \bottomrule
        \end{tabular}
    }
    \label{tab:dataset}
\end{table}
Based on above-mentioned unit tests, VerilogEval~\citep{liu2023verilogeval} proposes to use pass@k metric to further reflect the correctness of generated RTL codes:
\begin{equation}
    pass@k = \underset{problems}{\mathbb{E}} \left [ 1 - \frac{\binom{n-c}{k}}{\binom{n}{k}} \right ],
\end{equation}
where we generate $n \geq k$ samples per task in which $c \leq n$ samples pass the unit test.

Some prior works, including AutoChip~\citep{thakur2023autochip}, ChipGPT~\citep{chang2023chipgpt}, \cite{thakur2023benchmarking}, categorize the problem set into kinds of difficulty: e.g. easy, intermediate and hard.
Under different difficulty settings, they evaluate the generated RTL codes with syntax goal and functionality goal, design quality goal.



\subsection{HDL Verification and Analysis}
Another promising application of LLMs for EDA is to leverage LLMs to understand, analyse and summarize the input RTL codes.
Different from Sec.~\ref{sec:eval-gen-code} evaluating the generated codes through external tools, LLMs take input as RTL codes and user-specified queries, and provide responses to the user based on this input.
This approach allows users to directly interact with the LLMs, reaching out to the analysis of input codes.

To assist engineers in bug summarization and analysis, ChipNeMo~\citep{liu2023chipnemo} constructs a domain-specific SFT dataset based on NVIDIA's internal bug database, NVBugs. Considering bug descriptions could be too large for context windows, they replace long path names with short alias, and split the summarization tasks into an incremental task.

Besides bug summarization and analysis,  RTLFixer~\citep{tsai2023rtlfixer} proposes a paradigm for fixing erroneous Verilog codes directly.
Formulating an input prompt and followed by utilizing Retrieval Augmented Generation (RAG) and ReAct prompting mechanism~\citep{yao2022react}, the agent revises the erroneous Verilog codes. If syntax errors persist, error logs from the compiler as well as retrieved human guidance from the database are provided as feedback.

LLMs also demonstrate the capability in assertion checking, which is effective at finding intricate RTL bugs and security issues.
It is a popular verification technique, where the specification of design under test in coded into assertions or properties in hardware description language.
Each assertion will focus on verifying individual functionality and logic. Besides, it is also able to detect security vulnerabilities to defense attacks.
\cite{orenes2023using} proposes an iterative methodology based on formal property verification to generate SystemVerilog Assertions (SVA) from a given RTL module. They also integrate this assertion generation flow in open-source framework, AutoSVA~\citep{orenes2021autosva}.
\cite{kande2023llm} realizes SVA with a similar flow, including prompt construction, LLMs-based assertions generation and simulation. They also lexical tools to automatically fix minor mistakes based on qualitative analysis about 'common pitfalls' made by LLMs.

Considering the security problems in hardware designs, NSPG~\citep{meng2023unlocking} proposes a framework to identify essential properties from Intellectual Property (IP) cores documentations. They fine-tune a pre-trained LLM and a sequence classification model on datasets consisting of hardware documentations, where the latter is responsible to identify whether a sentence in documentation is a property or not, which contains essential information of operator behaviour and security. 
DIVAS~\citep{paria2023divas} proposes a LLM-based end-to-end framework for identify the CWEs for a given System-on-Chip (SoC) specification and employs a novel LLM-based technique to determine the relevant CWEs, which are finally converted into SystemVerilog Assertions using LLMs for verification. 
\citep{ahmad2023fixing} utilizes a detector tool to extract or directly get the location and CWE type of bug. These bug information, codes before the bug and buggy code in comments are combined as LLM prompt, assisting LLM to fix this hardware bug.

\section{Datasets of LLM4EDA}
Fine-tuning general-purpose LLMs with extensive and accurate domain-specific data would yield better performance.
Some works have evolved in constructing corresponding datasets through collecting existed data, including public design source codes, documentation, internal error logs, and bug summaries.
Simultaneously, other works generate synthesis dataset by querying existed general LLMs for circuits or design instructions.
In Table~\ref{tab:dataset}, we present some representative works that have explored the fine-tuning of LLMs utilizing domain-specific datasets.



\section{Backbones for LLM4EDA}
We demonstrate the tasks, backbone and fine-tuning techniques for LLMs applied in EDA in Table~\ref{tab:type-size-LLMs}.
Due to limitations in computational resources and available datasets, current efforts are primarily focused on utilizing APIs of existing LLMs.
For domain specific data, they either collect existed RTL codes~\citep{liu2023chipnemo, fu2023gpt4aigchip, thakur2023verigen, dehaerne2023deep, meng2023unlocking}, or generate synthesis designs from pre-trained general LLMs~\citep{liu2023rtlcoder, liu2023verilogeval}.

\begin{table}[!tb]
    \centering
    \caption{
    Type and size of corresponding LLMs selected in each method.
    Chat: assistant chatbot. Gen.: HDL and script generation. V\&A: HDL verification and analysis.
    }
    \adjustbox{width=1.0\linewidth}{

    \begin{tabular}{c|ccc|c|c|ccccccc}
         \toprule
         \multirow{2}{*}{\textbf{Method}} & \multicolumn{3}{c|}{\textbf{Task}} & \multirow{2}{*}{\textbf{Type}} & \multirow{2}{*}{\textbf{Size}} & \multirow{2}{*}{\textbf{Fine-tune}}\\
         ~ & \textbf{Chat} & \textbf{Gen.} & \textbf{V\&A} & ~ & ~ & ~ \\
        \midrule
        \midrule
        RTLCoder~\citep{liu2023rtlcoder} & & & \checkmark & Zephyr & 7 B & Quality Score \\
        VerilogEval~\citep{liu2023verilogeval} & &  \checkmark  & & CodeGen & 16 B & SFT \\
        \cite{ahmad2023fixing} & & \checkmark & & CodeGen & 16 B \\
        \cite{thakur2023benchmarking} & & \checkmark & & CodeGen & 16 B & AI21 Studio\\
        ChipNeMo~\citep{liu2023chipnemo} & \checkmark & \checkmark  &  \checkmark & LLaMA2 & 13 B & SFT \\
        ChatEDA~\citep{he2023chateda} &  & \checkmark  & & LLaMA2 & 70 B &  QLoRA \\
        \midrule
        ChipGPT~\citep{chang2023chipgpt} &  & \checkmark & & GPT-3.5 & &\\
        DIVAS~\citep{paria2023divas} &  & & \checkmark & GPT-3.5  & &\\
        \cite{schafer2023empirical} &  & \checkmark & & GPT-3.5  & &\\
        \cite{nair2023generating} &  & \checkmark & & GPT-3.5  & &\\
        \cite{du2023power} &  & \checkmark & \checkmark & GPT-3.5  & &\\
        \cite{kande2023llm} &  & & \checkmark & GPT-3.5  & &\\
        RTLFixer~\citep{tsai2023rtlfixer} &  & & \checkmark & GPT-3.5  & &\\
        RTLLM~\citep{lu2023rtllm} &  & \checkmark & & GPT-4  & &\\
        GPT4AIGChip~\citep{fu2023gpt4aigchip} &  & \checkmark & & GPT-4  & &\\
        AutoChip~\citep{thakur2023autochip} &  & \checkmark & & GPT-4  & &\\
        Chip-Chat~\citep{blocklove2023chip} &  & \checkmark & & GPT-4  & &\\
        VeriGen~\citep{thakur2023verigen} &  & \checkmark & &  GPT-4  & &\\
        \cite{orenes2023using} &  & & \checkmark & GPT-4  & &\\

        \bottomrule
    \end{tabular}
    
    }
    \label{tab:type-size-LLMs}
\end{table}

\section{Outlook}
\subsection{LLM in Logic Synthesis}
Logic synthesis transforms a high-level description of a design, e.g. Verilog, into an optimized gate-level representation. 
It mainly consists of three steps, namely pre-mapping optimization, technology mapping, and post-mapping optimization.
Extensive tuning of the synthesis optimization flow is required, including which optimizations to use, their corresponding arguments and the order of invocation.
This process can be treated as a sequential decision making problem in parameterized action space~\citep{fan2019hybrid}.
Efficient design space exploration is challenging due to the exponential number of possible optimization permutations.
Therefore, it heavily relies on experienced engineers.

Current works utilize heuristics~\citep{li2022himap}, Bayesian Optimization (BO)~\citep{grosnit2022boils} Reinforcement Learning (RL)~\citep{hosny2020drills, yuan2023easyso} to search the well-performed optimization sequence.
With respect to LLM, taking HDL of design and appropriate prompts as input (e.g., containing target PPA), LLM can generate the sequence of optimization along with their respective arguments.
Furthermore, final or estimated PPA can serve as the feedback for LLM, enabling iteratively enhancement of the Quality of Results (QoR).

\subsection{LLM in Physical Design}
\label{sec:LLM-in-Physical-Design}
In physical design, placement algorithm determines location for each module, including macro module and standard cell.
It aims to minimize wirelength cost subjecting to density constraints.
After placement, the (wire) routing step adds wires needed to properly connect the placed components while obeying all design rules.
The main objective is to connect all the required connections and on this basis, reduce the routing wirelength and overflow.
Placement and Routing (P\&R) consumes major part of time and computational resources during physical design, and has a significant impact on final PPA.

The utilization of LLMs in Placement and Routing (P\&R) remains relatively unexplored due to the inherent challenges posed by large-scale problems. 
Contemporary designs typically comprise millions of cells and nets~\citep{lin2019dreamplace}, making it impractical for LLMs to directly address the location and wire assignment of each component. 
Employing graph partitioning and clustering algorithms provides feasible approaches to reduce the problem's scale. 
While these methods may result in a loss of fine-grained details and potential constraint violations, the resulting `coarse' layouts can serve as initial solutions for traditional P\&R algorithms, thereby expediting the solving process.

An alternative strategy could involve optimizing the modules, which are few in number but subject to numerous complex constraints. In this scenario, LLMs could serve dual purposes: they could be employed to generate the optimal outcome, or they could act as a black-box reward system, providing feedback to the reinforcement learning agent.

\subsection{Feature Extraction and Alignment}
Currently, circuits are typically expressed using hardware description languages, representing them in a textual format. 
Once the synthesis process is completed, the circuit is transformed into an equivalent netlist, commonly represented as a directed acyclic graph (DAG). 
Subsequently, placement and routing operations are performed to generate a layout, establishing the physical positions and interconnections of components.

These representations stem from the same underlying entity and correspond to language-based, graph-based, and image-based representations, respectively. 
However, extraction and alignment of these multi-modal contents has not been thoroughly investigated to date. 
Large Graph Models~\citep{zhang2023large, tang2023graphgpt} and Large Vision Models~\citep{wang2023review, 2023GPTvision} have demonstrate strong potential to extract informative feature and representation from netlist and layout respectively.
We could also build a large-scale multi-modal pre-trained model~\citep{wang2023large} to realize alignment of there features from different modalities.
Techniques such as contrastive learning, e.g. CLIP (Contrastive Language-Image Pretraining)~\citep{radford2021learning}, or methods involving masking and reconstructing some modalities from others, can be employed to achieve content alignment.

\subsection{Long Chain Feedback for PPA}
With the assistance of LLMs, our aim is to improve the PPA in chip design. 
However, there is still a significant gap in predicting and optimizing PPA at the system design stage due to the necessity of numerous intermediate processes such as synthesis, placement, and routing. 
This lengthy feedback chain poses challenges in optimizing PPA from the initial stages.
We believe this problem could be addressed from following perspectives:
\begin{enumerate}[leftmargin=0.175in]
    \item \textbf{Domain specific datasets with PPA.} 
    Domain-specific datasets incorporating PPA metrics have garnered attention. 
    Existing datasets utilized for fine-tuning general pre-trained LLMs typically include source codes and corresponding descriptions.
    However, they lack awareness of PPA considerations~\citep{chang2023chipgpt}. 
    By augmenting these datasets with final PPA metrics, they can serve as more accurate design specifications or targets, as well as providing supervised labels. 
    This augmentation would bridge the gap between RTL designs and their corresponding final PPA, enhancing the effectiveness and applicability of the datasets for fune-tuning PPA-aware LLMs.
    
    \item \textbf{Utilize PPA as feedback.} 
    For a human engineer, PPA is an essential feedback serving as guidance in the iterative and repetitive design process.
    Similarly, during the interaction with LLMs, we can either 1) evaluate the final PPA with high accuracy through external verification tools or 2) predict PPA with enhanced computational speed, e.g. with neural network based methods~\citep{guo2022timing, yang2022versatile, zhong2024preroutgnn}.
    These PPA could serve as feedback for LLMs, guiding them to generate or refine designs with improved quality.
    
    \item \textbf{Prompt engineering and task planning.}
    Prompt matters in interacting with LLMs especially for querying in EDA field~\citep{chang2023chipgpt, he2023chateda, pearce2020dave}.
    We could construct more informative prompts by specifying PPA targets, incorporating cost constraints, and integrating design rules, facilitating LLMs to better understand and realize our purposes.
    To make full use of the context window, it is beneficial to break down large design through task planning into sub-tasks, where each has its own interface communicating with others.
    Additionally, the output generated from previous tasks could serve as a fundamental specification and guidance for subsequent sub-tasking processing, thus facilitating a coherent and efficient design workflow.
    
\end{enumerate}


\section{Conclusion}
This paper presents a comprehensive survey on the integration of Language Models (LLMs) in the Electronic Design Automation (EDA) field. 
The survey encompasses a range of applications of LLMs in EDA, namely: 1) assistant chatbot, 2) generation of HDL code and EDA flow scripts, 
3) verification and analysis of HDL code. 
Additionally, we highlight the future research direction, focusing on applying LLMs in logic synthesis and physical design, addressing long feedback of PPA, multi-modal feature extraction and alignment of circuits.
We firmly anticipate the eventual realization of large-scale multi-modal pre-trained models, for but not only for EDA.

\clearpage
\bibliographystyle{plainnat}
\bibliography{reference}

\end{document}